\documentclass[prd,aps,tightenlines,superscriptaddress,amsmath,amssymb,amsfonts,twocolumn]{revtex4-2}
\usepackage{amssymb,amsmath,amsthm,amsbsy,epsfig,color,graphicx,times}
\usepackage[ansinew]{inputenc}
\usepackage[english]{babel}
\usepackage{color}
\usepackage{braket}
\usepackage{tabularx}
\usepackage{array}

\begin{document}
\title{Boson mixing and flavor oscillations in curved space-time}

\author{A. Capolupo}
\email{capolupo@sa.infn.it}
\affiliation{Dipartimento di Fisica ``E.R. Caianiello'' Universit\`{a} di Salerno, and INFN -- Gruppo Collegato di Salerno, Via Giovanni Paolo II, 132, 84084 Fisciano (SA), Italy}

\author{A. Quaranta}
\email{anquaranta@unisa.it}
\affiliation{Dipartimento di Fisica ``E.R. Caianiello'' Universit\`{a} di Salerno, and INFN -- Gruppo Collegato di Salerno, Via Giovanni Paolo II, 132, 84084 Fisciano (SA), Italy}

\author{P. A. Setaro}
\email{p.setaro@studenti.unisa.it}
\affiliation{Dipartimento di Fisica ``E.R. Caianiello'' Universit\`{a} di Salerno, Via Giovanni Paolo II, 132, 84084 Fisciano (SA), Italy}

\begin{abstract}

We develop a quantum field theory of boson mixing in curved space. We derive new general oscillation probabilities and prove that the formalism correctly reproduces the flat space limit. We explicitly compute the oscillation formulae for two cosmologically relevant Friedmann-Lemaitre-Robertson-Walker metrics.

\end{abstract}

\maketitle

\maketitle

\section{Introduction}

Among the known elementary particles only neutrinos undergo the phenomenon of flavor oscillations \cite{Neutrino1,Neutrino2}. The quantum field theory (QFT) of fermion mixing, which is adequate to the description of mixed neutrinos, shows a very rich structure both in flat \cite{FMix3,FMix1,FMix2,Fmix4} and in curved spacetime \cite{Capolupo2020}. In virtue of the peculiar condensate structure of the flavor vacuum, this theory may have interesting cosmological consequences \cite{FDM,FDM1,FDM2,FDM3,FDM4}. Simulations of the QFT of flavor mixing, in the context of atomic physics, have also been proposed \cite{Sim}. The mixing of bosons is, in some respects, more exotic. The bosons which are known to mix are indeed composite particles subject to decay, i. e. the mesons $K, B, D$ \cite{Kabir,GellMann,Jubb,Bitenc,Meson1,Meson2,Meson3,Meson4,Meson5,Meson6,Meson7,Meson8,Meson9,Meson10,Meson11}. Other schemes of boson mixing may arise under specific conditions for hypothetical particles, as it is the case of the mixing of photons with axionlike particles in presence of an external magnetic field \cite{Axion1,Axion2}. One can additionally speculate that the supersymmetryc bosonic partners of neutrinos \cite{sneutrino1,sneutrino2} would mix in analogy with their fermionic counterparts. The QFT of boson mixing, developed in flat space in  (\cite{BMix,BMix2} and references therein) is then interesting in order to provide a solid field theoretical foundation to the phenomena mentioned above. Just like the QFT of fermion mixing, the QFT of boson mixing presents a rich mathematical structure and a nontrivial vacuum state. Previous analyses have shown that the latter may contribute to the dark components of the universe, in a clear analogy with the fermion flavor vacuum \cite{FDM1,FDM2}.

In this paper we wish to construct the QFT of boson mixing in curved spacetime, generalizing the flat space theory. Such a step is not only essential to analyze the boson oscillations in curved space, but also to provide an adequate basis for the study of the bosonic flavor vacuum in a cosmological context. The formalism presented here may indeed substantiate the connection between the bosonic flavor vacuum and dark energy, which was outlined in previous studies \cite{FDM1,FDM2} on the basis of the flat space theory.

Our construction closely parallels, mutatis mutandis, the QFT of fermion mixing in curved space developed in the ref. \cite{Capolupo2020}. Adjoining field mixing and curved space quantization results in a complex scenario, where the existence of infinitely many unitarily inequivalent representations of the canonical commutation relations is not only due to the mixing transformations, but also due to the intrinsic ambiguity underlying the field quantization on curved space \cite{Curv1,Curv2,Curv3,Curv4}. The interplay between mixing and curvature does indeed constitute the core of our analysis.
We first construct the QFT of boson mixing in a fixed but arbitrary representation of the free fields, by appropriately generalizing the flat space procedure. We define the oscillation probabilities from the Noether charges of the Lagrangian. Then we study the transformations induced by a change in the representation of the free fields. We exhibit the invariance of local observables on the one hand and the partial invariance of oscillation probabilites on the other hand. We argue that only under specific compatibility conditions the probabilities are left invariant by a change in the representation of the free fields. Then we apply the formalism to two cosmological metrics, deriving explicitly new oscillation formulae.

The paper is structured as follows. In section II we recall the basics of free field quantization in curved space and set the notation. In section III we generalize the mixing transformations to curved space and apply them to the fields with definite masses. We introduce the oscillation probabilities and give an explicit expression in terms of the Bogoliubov coefficients of the mixing transformations. In section IV we deal with changes of representation of the free fields, and we determine under which conditions the resulting probabilities are invariant. We then apply the formalism to some specific spacetimes. In section V we show that the flat spacetime limit is correctly reproduced. In section VI we work out the oscillation formulae in two spatially flat Friedmann-Lemaitre-Robertson-Walker spacetimes, corresponding respectively to a universe dominated by the cosmological constant and by radiation. Section VII is devoted to conclusions.

\section{Free bosons in curved space-time}

In order to evaluate the oscillation formulae for bosons on a curved space-time, it is necessary to consider both the effects of curvature and mixing on the (free) mass fields.
The boson fields with definite masses satisfy the (free) Klein-Gordon equations in curved space
\begin{equation}\label{KGequation1}
 \frac{1}{\sqrt{-g}} \partial_{\mu} \left(\sqrt{-g} g^{\mu \nu} \partial_{\nu} \phi_i \right) + m^2_i \phi_i = 0
\end{equation}

We shall assume that the bosons carry some kind of $U(1)$ charge (e.g. ``baryon number'', ``strangeness'', etc.), so that particles and antiparticles are distinct. Due to the linearity of eq. \eqref{KGequation1}, the free boson fields can  be expanded as

\begin{equation}\label{FreeFieldExpansion}
 \phi_i (x)=\sum_{k} \left( \gamma_{k;i} \zeta_{k;i} (x) + \epsilon_{k;i}^{\dagger} \zeta^{*}_{k;i} (x) \right)
\end{equation}

where $k$ is a generic mode index. For ease of treatment with general metrics all the spacetime dependency of the field is condensed within the modes. The meaning of $k$ depends on the underlying metric. For metrics that display a spatial translational invariance, such as the Minkowski or the spatially flat Friedmann-Robertson-Walker metrics, the natural choice is the three-momentum $\pmb{k}$.
The operator coefficients $\gamma_{k,i}, \epsilon_{k,i}$ are the annihilators satisfying the usual canonical commutation relations, while $\zeta_{k,i}$ and $\zeta^{*}_{k,i}$ are a set of positive and negative frequency, solutions of the Klein-Gordon equation with mass $m_i$.
Operators with distinct mass indices $i \neq j$ are assumed to commute, so that $[\gamma_{k,i}, \gamma^{\dagger}_{k,j}] = \delta_{ij}$, $[\epsilon_{k,i}, \epsilon^{\dagger}_{k,j}] = \delta_{ij}$. A specific choice of the positive frequency modes $\zeta_{k,i}$, possibly with respect to some specified observer, must be made in order to construct the canonical theory and develop a particle interpretation. This choice defines the Fock space $H_{m} = H_1 \otimes H_2$ and the corresponding vacuum state as  $|0_{m} \rangle = |0_1 \rangle \otimes |0_2 \rangle$, where $\ket{0_i}$, with $i=1,2$, is defined as usual by  $\gamma_{k,i}\ket{0_i} = 0 = \epsilon_{k,i}\ket{0_i} $ for each $k,i$.

Of course the choice of the positive frequency modes is not unique and any other arbitrary basis $\{\tilde{\zeta}_{k,i},\tilde{\zeta^{*}}_{k,i}\}$ can be used to expand the fields, i.e., $\phi_i $ can be written also as $\phi_i = \sum_{k}( \tilde{\gamma}_{k,i} \tilde{\zeta}_{k,i} (x) + \tilde{\epsilon}_{k,i}^{\dagger} \tilde{\zeta^{*}}_{k,i} (x))$.
Since both the sets $\{\zeta_{k,i},\zeta^{*}_{k,i}\}$ and $\{\tilde{\zeta}_{k,i},\tilde{\zeta^{*}}_{k,i}\}$ form a basis for the space of the solutions of the Klein-Gordon equation, one can write the modes of a set in terms of the modes of the other set, and for any $i$, one has:
\begin{eqnarray}\label{BogoliubovModes}
 \nonumber \tilde{\zeta}_{k',i} &=& \sum_{k} \left( \Gamma_{k';k;i} \zeta_{k,i} + \Sigma_{k';k; i}\zeta^{*}_{k, i} \right) \\
  \tilde{\zeta}^{*}_{k', i} &=& \sum_{k} \left( \Gamma^{*}_{k';k; i} \zeta^{*}_{k,i} + \Sigma^{*}_{k';k;i}\zeta_{k,i} \right)\,,
\end{eqnarray}
where, $ \Gamma_{k'; k; i} = ( \tilde{\zeta}_{k',i}, \zeta_{k,i})$ and $\Sigma_{k';k;i} =( \tilde{\zeta}_{k',i},\zeta^{*}_{k,i})$ are  Bogoliubov coefficients, such that \mbox{$\sum_{q} \left( \Gamma_{k;q;i}^{*} \Gamma_{k';q;i} - \Sigma_{k;q;i}^{*} \Sigma_{k';q;i} \right) = \delta_{k,k'}$} for any $i$.

% Note that if $\zeta_{k',i}$ are positive frequency modes with respect to same timelike Killing vector field $\tilde{\zeta}_{k',i}$ are a linear combination $\zeta_{k',i}$ alone, conteining only positive frequency respect Killing vector and two set of modes share a common vacuum state.

The corresponding relations between the two sets of annihilators is the following
\begin{eqnarray}\label{Bogoliubov1}
% \nonumber % Remove numbering (before each equation)
 \nonumber \tilde{\gamma}_{k,i} &=& \sum_{k'} \left(\Gamma^{*}_{k; k';i} \gamma_{k',i} -  \Sigma^{*}_{k;k';i} \epsilon^{\dagger}_{k',i} \right)\\
  \tilde{\epsilon}_{k,i} &=& \sum_{k'} \left( \Gamma^{*}_{k; k';i} \epsilon_{k',i} -  \Sigma^{*}_{k;k';i} \gamma^{\dagger}_{k',i} \right) \ .
\end{eqnarray}

In many cases the Bogoliubov coefficients $\Gamma_{k; k';i} , \Sigma_{k; k';i}$ can be expressed as $\Gamma_{k; k';i} = \delta_{k,k'}\Gamma_{k,i}$, $\Sigma_{k; k';i} = \delta_{k,k'} \Sigma_{k,i}$, with $\Gamma_{k,i}$ and $\Sigma_{k,i}$ depending only on $k$.
Moreover, the Bogoliubov transformations in Eqs. \eqref{Bogoliubov1} can be recast in terms of the generators
\begin{eqnarray}
J_i = e^{\sum_{k,k'} \left[ ( \lambda_{k,k',i}^* \gamma_{k,i}^{\dagger}\epsilon_{k',i}^{\dagger} + \lambda_{k,k',i} \epsilon_{k,i} \gamma_{k',i})\right] } ,
\end{eqnarray}
 with  $\lambda_{k,k',i} = Arctan(\frac{\Sigma_{k;k';i}}{\Gamma_{k;k';i}})$, as
$
  \tilde{\gamma}_{k,i} = J_i^{-1} \gamma_{k,i} J_i  $  and $  \tilde{\epsilon}_{k,i} = J_i^{-1} \epsilon_{k,i} J_i $.

The generators $J_i$ map the Fock spaces $\mathcal{H}_i$ built from the $\gamma_{k,i},\epsilon_{k,i}$ into the Fock spaces $\tilde{\mathcal{H}}_i$ built from the $\tilde{\gamma}_{k,i},\tilde{\epsilon}_{k,i}$,  $J_i:\tilde{\mathcal{H}}_i \rightarrow \mathcal{H}_i $ .
In particular, one has for the vacuum states $|\tilde{0}_i\rangle = J_i^{-1} |0_i \rangle$.
Similarly to the untilded representation, the mass Hilbert space in the tilded representation is given by  a tensor product: $\tilde{\mathcal{H}}_m = \tilde{\mathcal{H}}_1 \otimes \tilde{\mathcal{H}}_2$. It is also convenient to define a unique generator of the Bogoliubov transformations $J: \tilde{\mathcal{H}}_m \longrightarrow \mathcal{H}_m$ on $\tilde{\mathcal{H}}_m$ as the tensor product $J = J_1 \otimes J_2$. Then, by definition
\begin{equation}\label{CurvatureGen2}
  \tilde{\gamma}_{k,i} = J^{-1} \gamma_{k,i} J \;,  \qquad  \qquad \ \ \tilde{\epsilon}_{k,i} = J^{-1} \epsilon_{k,i} J
\end{equation}
for $i = 1,2$.

\section{Boson mixing and oscillations in curved space-time}

The mixing relations, for two flavor fields are given by
\begin{eqnarray}\nonumber\label{mix}
\phi_{A} = \cos(\theta) \phi_1 +  \sin(\theta) \phi_2
\\ \phi_{B} = \cos(\theta)\phi_2 - \sin(\theta) \phi_1
 \end{eqnarray}
 where $\phi_{A}$ and $\phi_{B}$ are the fields with definite flavors, $\phi_{1}$ and $\phi_{2}$ are the fields with definite mass, and $\theta$ is the (2-flavor) mixing angle.
Introducing the mixing generator $\mathcal{G}_{\theta} (\tau)$, Eqs.(\ref{mix}) can be written as
 \begin{equation}\label{GFields}\nonumber
\phi_A = \mathcal{G}_{\theta}^{-1} (\tau) \phi_1 \mathcal{G}_{\theta} (\tau) \;,
 \\
 \phi_{B} = \mathcal{G}_{\theta}^{-1} (\tau) \phi_2 \mathcal{G}_{\theta} (\tau)\,,
\end{equation}
where $\mathcal{G}_{\theta} (\tau)$ is
\begin{equation}\label{CurvedMixingGenerator}
  \mathcal{G}_{\theta} (\tau) =  e^{\theta [(\phi_1,\phi_2)_{\tau}-(\phi_2,\phi_1)_{\tau}]} \ .
\end{equation}
Here the scalar products $(\phi_i,\phi_j)_{\tau}$ \emph{do} depend on the hypersurface chosen for the integration, since they are solutions to different Klein-Gordon equations. The Klein-Gordon inner product is defined, as usual, as

\begin{equation}\label{InnerProduct}
 (u,v)_{\tau} = -i \int_{\Sigma_{\tau}} d \Sigma^{\mu} \sqrt{-g} \left(A^{*} \partial_{\mu} B - B\partial_{\mu}A^{*}\right) \ ,
\end{equation}

where the parameter $\tau$ labels a foliation by Cauchy hypersurfaces and it is understood that the underlying spacetime is globally hyperbolic. The integral is to be performed over the hypersurface $\Sigma_{\tau}$ and $g = \det \left( g_{\mu \nu}\right)$. If $u$ and $v$ are solutions to the same Klein-Gordon equation the $\tau$ dependence disappears, while, as anticipated, the inner product depends on $\tau$ if $u$ and $v$ are solutions to distinct Klein-Gordon equations.

For any mass representation, we can build the flavor spaces $\mathcal{H}_{f}(\tau),$ by means of the mixing generator $\mathcal{G}_{\theta} (\tau) : \mathcal{H}_m (\tau) \rightarrow \mathcal{H}_f $, and when the generator \eqref{CurvedMixingGenerator} act on the mass annihilators, one obtains the flavor annihilators for  curved space

\begin{widetext}
\begin{equation}
  \gamma_{k,A} (\tau) =  \mathcal{G}_{\theta}^{-1} (\tau) \gamma_{k,1} \mathcal{G}_{\theta} (\tau)=  \cos(\theta) \gamma_{k, 1} + \sin(\theta)    \sum_{q } \bigg[\Lambda^*_{q, k }(\tau) \gamma_{q, 2}   + \Xi_{q, k }(\tau) \epsilon_{q, 2}^{\dagger}\bigg]\,.
\end{equation}
\end{widetext}
Similar equations hold for $\gamma_{k,B} (\tau), \epsilon_{k,A} (\tau), \epsilon_{k,B} (\tau)$.
The Bogoliubov coefficients are given by the inner products of the solutions of the \emph{curved space} Klein-Gordon equation with mass $m_1$ and $m_2$: $\Lambda_{q;k} (\tau) = (\zeta_{q,2},\zeta_{k,1})_{\tau}$ and $\Xi_{q;k} (\tau) = (\zeta_{k,1},\zeta^{*}_{q,2})_{\tau}$ and satisfy the condition

\begin{equation}\label{FermionBog}
\sum_{q} \left(\Lambda_{k;q}^{*}(\tau) \Lambda_{k';q}(\tau)-\Xi_{k;q}^{*}(\tau) \Xi_{k';q} (\tau) \right)\! = \! \delta_{k,k'} \ ,
\end{equation}

for each $\tau$.
The mass and flavor representations are unitarily inequivalent in the infinite volume limit. At the finite volume, the flavor vacuum is connected to the mass vacuum by means of the generator  $\mathcal{G}_{\theta}$ as $\ket{0_f (\tau)} = \mathcal{G}_{\theta}^{-1} (\tau) \ket{0_m}$. It is a condensate of particle-antiparticle pairs with defined mass and opposite momentum.
Moreover, for any $\tau$, one has a distinct flavor Fock space $\mathcal{H}_f (\eta)$ defined by the operators $\gamma_{A,B} (\tau) , \epsilon_{A,B} (\eta)$.

Apart from particular expansions, as is the case when the mass fields are expanded in terms of modes labelled by the energy,  the mixing Bogoliubov coefficients are  diagonal, so that they can be written as:
\begin{eqnarray}\label{Compatibility}
 \nonumber \Lambda_{q,k} (\tau) &=& \delta_{q,k}  \Lambda_{k }(\tau) \\
 \Xi_{q, k } (\tau) &=& \delta_{q,k}  \Xi_{k }(\tau)
\end{eqnarray}
with $\Lambda_{k }(\tau)$, $\Xi_{k }(\tau)$ depending on $k$   alone.

Cleary the total Lagrangian, which is simply the sum of two free Klein-Gordon Lagrangians with masses $m_1$ and $m_2$, is invariant under global $U(1)$ gauge transformations. This implies that the total charge $Q = Q_1 + Q_2 = Q_A + Q_{B}$ is conserved, where $Q_i =\sum_{k} Q_{i}^{k}= \sum_{k } \left( \gamma_{k, i}^{\dagger} \gamma_{k, i} - \epsilon_{k, i}^{\dagger} \epsilon_{k, i} \right)$ for $i=1,2$ or $A,B$. The transition probabilities can be then defined as

\begin{widetext}
\begin{equation}\label{TransitionProb}
   P^{\rho \rightarrow \sigma}_{k} (\tau) = \sum_{q} \bigg(\langle \phi_{\rho,k} (\tau_0) | Q^{q}_{\sigma} (\tau) | \phi_{\rho,k} (\tau_0) \rangle
 - \langle 0_f(\tau_0) | Q^{q}_{\sigma} (\tau) |0_f (\tau_0) \rangle \bigg).
\end{equation}
\end{widetext}
Here $\rho,\sigma = A,B$,  and  $| \phi_{\rho,k} (\tau_0) \rangle = \gamma^{\dagger}_{k,\rho} (\tau_0) \ket{0_f(\tau_0)}$ is the state with a single particle of flavor $\rho$, quantum numbers $k$ on the reference hypersurface $\tau = \tau_0$.
The second term on the rhs of Eq.\eqref{TransitionProb} represents the normal ordering with respect to $\ket{0_f(\tau_0)}$.
It is possible to identify the expectation values of charge operator with oscillation probabilities since  the sum of expectations values corresponding to all possible transitions is constant and equal to $1$:
  $ P^{A \rightarrow A}_{k}(\tau) +   P^{A \rightarrow B}_{k}(\tau) = 1$ and $P^{B \rightarrow A}_{k}(\tau) +   P^{B \rightarrow B}_{k}(\tau) = 1$ for any $\tau$.

It is straightforward to see that in terms of flavor operators, Eq.\eqref{TransitionProb} can be written as
\begin{widetext}
\begin{equation}\label{ProbCommutator}
 P^{A \rightarrow B}_{k}(\tau)=\sum_{k'}\left(|[\gamma_{k',B}(\tau),\gamma^{\dagger}_{k,A}(\tau_{0})]|^{2}-|[\epsilon^{\dagger}_{k',B}(\tau),\gamma^{\dagger}_{k,A}
 (\tau_{0})]|^{2}\right)\,.
\end{equation}
By computing explicitly the commutators in Eq.(\ref{ProbCommutator}), we obtain the oscillation formula
\begin{equation}\label{DefinitiveProbs01}
  P^{A \rightarrow B}_{k}(\tau)  =   2 \cos^{2}(\theta) \sin^{2}(\theta)
   \bigg[1-\sum_{q} \Re\bigg(\Lambda_{k;q}^{*} (\tau_0) \Lambda_{k;q} (\tau)
  -\Xi^{*}_{k;q} (\tau_0) \Xi_{k;q} (\tau)  \bigg)\bigg] \ ,
\end{equation}
which represents the main result of the paper.
When Eqs.\eqref{Compatibility} hold, Eq.(\ref{DefinitiveProbs01}) reduces to
\begin{equation}\label{DefinitiveProbs}
   P^{A \rightarrow B}_{k}(\tau)  =   2 \cos^{2}(\theta) \sin^{2}(\theta)
    \bigg[1 -\Re\bigg(\Lambda_{k}^{*} (\tau_0) \Lambda_{k} (\tau)
  -  \Xi_{k}^{*} (\tau_0) \Xi_{k} (\tau)  \bigg)\bigg] \ .
\end{equation}
Notice the striking similarity, apart from a relative sign, of eq. \eqref{DefinitiveProbs01} with the analogous probability for fermions derived in the ref. \cite{Capolupo2020}.

\end{widetext}

\section{Ambiguity in particle interpretation and invarance of the transition probabilities}

In deriving eqs. \eqref{DefinitiveProbs01} and \eqref{DefinitiveProbs} we have assume a fixed, albeit arbitrary, representation of the mass fields. We need now to determine how the flavor operators and the transition probabilities vary when the mass representation is changed.

The mass Fock spaces of the representations $\{\gamma_1,\epsilon_1 \}, \{\gamma_{2}, \epsilon_2 \}$, and $\{\tilde{\gamma}_1,\tilde{\epsilon}_1 \}, \{ \tilde{\gamma}_2, \tilde{\epsilon}_2 \}$ are connected by he generator of eq. \eqref{CurvatureGen2} $J^{-1}:\mathcal{H}_m \rightarrow \tilde{\mathcal{H}}_{m}$.
We   build the corresponding flavor annihilators and flavor spaces $\mathcal{H}_{f}(\tau),\tilde{\mathcal{H}}_f (\tau)$, and the mixing generators $\mathcal{G}_{\theta} (\tau) : \mathcal{H}_f (\tau) \rightarrow \mathcal{H}_m $, $\tilde{\mathcal{G}}_{\theta}(\tau): \tilde{\mathcal{H}}_f (\tau) \rightarrow \tilde{\mathcal{H}}_m$.
Moreover, we derive the relations among the mixing coefficients $\Lambda(\tau), \Xi (\tau) $ and $\tilde{\Lambda}(\tau), \tilde{\Xi}(\tau) $ that appear in the  generators $\mathcal{G}_{\theta} (\tau)$ and $\tilde{\mathcal{G}}_{\theta} (\tau)$.
The definition of   $\tilde{\Lambda} (\tau)$ is
\begin{equation}\label{TildeCoeff}
  \tilde{\Lambda}_{q;k}(\tau) = (\tilde{\zeta}_{q;2}, \tilde{\zeta}_{k;1})_{\tau}
  \end{equation}
 By using Eqs.\eqref{Bogoliubov1}, one has
 \begin{widetext}

 \begin{equation}\label{TildeCoeff}
  \tilde{\Lambda}_{q;k}(\tau) = \sum_{q',k'}\bigg(\bigg[\Gamma_{q;q';2} \zeta_{q';2} + \Sigma_{q;q';2} \zeta^{*}_{q';2}\bigg]  , \bigg[\Gamma_{k;k';1} \zeta_{k';1} + \Sigma_{k;k';1} \zeta^{*}_{k';1}\bigg]\bigg)_{\tau}\,.
 \end{equation}
Taking into account the properties of the inner product \eqref{InnerProduct}, in the general case we have
  \begin{eqnarray}
  \nonumber   \tilde{\Lambda}_{q;k}(\tau) &=& \sum_{q',k'} \bigg[ \Gamma^{*}_{q;q';2} \Gamma_{k;k';1} (\zeta_{q';2},\zeta_{k';1})_{\tau} + \ \ \Gamma^{*}_{q;q';2} \Sigma_{k;k';1} (\zeta_{q',r';2},\zeta^{*}_{k';1})_{\tau} \\  &+& \ \ \Sigma^{*}_{q;q';2} \Gamma_{k;k';1} (\zeta^{*}_{q';2},\zeta_{k';1})_{\tau}  + \ \ \Sigma^{*}_{q;q';2} \Sigma_{k;k';1} (\zeta^{*}_{q';2},\zeta^{*}_{k';1})_{\tau} \bigg]\,,
  \end{eqnarray}
and, using the definitions of the Bogoliubov coefficients $\Lambda(\tau)$ and $\Xi(\tau)$, we have
  \begin{eqnarray}  \label{MixingCoefficient1}
     \tilde{\Lambda}_{q;k}(\tau) =\sum_{q',k'} \bigg[ \Gamma^{*}_{q,q';2}
     \left(\Gamma_{k;k';1} \Lambda_{q';k'}(\tau)-  \Sigma_{k;k';1} \Xi_{q';k'}(\tau) \right)
   + \ \Sigma^{*}_{q;q';2}
    \left( \Gamma_{k;k';1} \Xi_{q';k'}^{*} (\tau)  -  \Sigma_{k;k';1} \Lambda_{q';k'}^{*}(\tau) \right) \bigg]
   \ .
\end{eqnarray}
 Similarly, we have
\begin{eqnarray}\label{MixingCoefficient2}
% \nonumber % Remove numbering (before each equation)
  \tilde{\Xi}_{q;k}(\tau) =  \sum_{q',k'} \bigg[ \Gamma^{*}_{q;q';2}
 \left(\Gamma^{*}_{k;k';1} \Xi_{q';k'}(\tau) -   \Sigma^{*}_{k;k';1} \Lambda_{q';k'}(\tau) \right)
   +  \ \Sigma^{*}_{q;q';2}
  \left( \Gamma^{*}_{k;k';1} \Lambda_{q';k'}^{*} (\tau) - \ \ \Sigma^{*}_{k;k';1} \Xi_{q';k'}^{*}(\tau) \right)\bigg] \
 .
 \end{eqnarray}

 \end{widetext}
When Bogoliubov coefficients are diagonal, Eqs.(\ref{MixingCoefficient1}) and (\ref{MixingCoefficient2}) reduce to
  \begin{eqnarray}\label{MixingCoefficient3}
  \nonumber  \tilde{\Lambda}_{q}(\tau) &=&   \sum_{q'} \bigg[ \Gamma^{*}_{q,q';2}
     \left( \Gamma_{q;q';1} \Lambda_{q'}(\tau) -  \Sigma_{q;q';1} \Xi_{q'}(\tau) \right) \\
    &+& \ \Sigma^{*}_{q;q';2}
    \left(\Gamma_{q;q';1} \Xi_{q'}^{*} (\tau) -  \Sigma_{q;q';1} \Lambda_{q'}^{*}(\tau) \right)\bigg]
    ,
\end{eqnarray}
and
\begin{eqnarray}\label{MixingCoefficient4}
% \nonumber % Remove numbering (before each equation)
 \nonumber  \tilde{\Xi}_{q}(\tau) &=&  \sum_{q'} \bigg[ \Gamma^{*}_{q;q';2}
  \left( \Gamma^{*}_{q;q';1} \Xi_{q'}(\tau) -   \Sigma^{*}_{q;q';1} \Lambda_{q'}(\tau) \right)\\
 &+& \   \Sigma^{*}_{q;q';2} \left( \Gamma^{*}_{q;q';1} \Lambda_{q'}^{*} (\tau)  -   \Sigma^{*}_{q;q';1} \Xi_{q'}^{*}(\tau) \right) \bigg]  \ .
\end{eqnarray}
Eqs. (\ref{MixingCoefficient1}), (\ref{MixingCoefficient2}, (\ref{MixingCoefficient3}), and  (\ref{MixingCoefficient4}) represent the relations existing among the mixing coefficients corresponding to the tilded and untilded mass representation.

The flavor vacuum $|0_f (\tau) \rangle$, under a change of mass representation, transforms as
\begin{equation}\label{FlavorTransform}
 |\tilde{0}_f (\tau) \rangle  = J_f^{-1} (\tau) |0_f (\tau) \rangle \doteq \tilde{\mathcal{I}}_{\theta}^{-1} (\tau) J^{-1} \mathcal{I}_{\theta} (\tau) |0_f (\tau) \rangle \
\end{equation}
where  $J_f^{-1} (\tau):\mathcal{H}_{m} \rightarrow \tilde{\mathcal{H}}_{m}$. In order that local observables be independent of the representation, the flavor operators must transform as
\begin{eqnarray}\label{OpFlavorTransform}
\nonumber \gamma_{k,\rho}(\tau) &\rightarrow& J_f^{-1}(\tau)  \gamma_{k,\rho}(\tau) J_f (\tau)  \\ \epsilon_{k,\rho}(\tau) &\rightarrow& J_f^{-1}(\tau)  \epsilon_{k,\rho}(\tau) J_f (\tau)\,.
\end{eqnarray}
Similar relations hold for the creation operators.

Equations \eqref{FlavorTransform} and \eqref{OpFlavorTransform} do indeed ensure that the local observables are independent of the mass representation. In general this is not the case for the transition probabilities, since they are not, strictly speaking, local observables.
For this reason they are invariant only under a restricted set of transformations, connecting mass representations that refer to the same kind of particle and therefore agree on the meaning of the quantum numbers $k$. Such representations are connected by diagonal Bogoliubov transformations
\begin{eqnarray}\label{BogoliubovModes2}
 \nonumber \tilde{\zeta}_{k,i} &=&  \Gamma_{k,i}^{*} \zeta_{k,i} + \Sigma_{k,  i}^{*} \xi_{k, i} \\
  \tilde{\xi}_{k, i} &=&  \Gamma_{k,  i} \xi_{k,i} - \Sigma_{k, i} \zeta_{k,i} \
\end{eqnarray}
where $\Gamma_{k;q;i} = \delta_{k,q} \Gamma_{k;i}$ and $\Sigma_{k;q;i} = \delta_{k,q} \Sigma_{k;i}$.

In order to show that the transition probabilities $P^{\rho \rightarrow \sigma}_{k}$ are invariant in this case, consider the tilded mass representation, with the transition probabilities given by (see eq. \eqref{DefinitiveProbs})

\begin{widetext}
\begin{eqnarray}\label{DefinitiveProbs2}
  \tilde{P}^{A \rightarrow B}_{k}(\tau)  =   2 \cos^{2}(\theta) \sin^{2}(\theta)
  \times  \bigg[1 - \sum_{q} \Re\bigg(\tilde{\Lambda}_{k;q}^{*} (\tau_0) \tilde{\Lambda}_{k;q} (\tau)
   -  \tilde{\Xi}_{k;q}^{*} (\tau_0) \tilde{\Xi}_{k;q} (\tau)  \bigg)\bigg]\,.
\end{eqnarray}
 Eqs. \eqref{DefinitiveProbs}  and  \eqref{DefinitiveProbs2} are manifestly equivalent if it holds the following equality
\begin{eqnarray}\label{PartialInvariance}
\Re[\tilde{\Lambda}_{k;q}^{*} (\tau_0) \tilde{\Lambda}_{k;q} (\tau) - \tilde{\Xi}_{k;q}^{*} (\tau_0) \tilde{\Xi}_{k;q} (\tau)] =
   \Re[\Lambda_{k;q}^{*} (\eta_0) \Lambda_{k;q} (\tau) -  \Xi_{k;q}^{*} (\eta_0) \Xi_{k;q} (\tau)] \,.
\end{eqnarray}
The equality can be proven by direct calculation
by using  Eqs.\eqref{MixingCoefficient1} and \eqref{MixingCoefficient2}
\begin{eqnarray}\label{ProductBogolibovCoeff}
\nonumber && \tilde{\Lambda}_{k;q}^{*} (\tau_0) \tilde{\Lambda}_{k;q} (\tau) -  \tilde{\Xi}_{k;q}^{*} (\tau_0) \tilde{\Xi}_{k;q} (\tau) = \\
 \nonumber && + \left (\Lambda_{k;q}^{*} (\tau_0) \Lambda_{k;q} (\tau) - \Xi_{k;q}^{*} (\tau_0) \Xi_{k;q} (\tau) \right)|\Gamma_{k,2}|^2
  \left[|\Gamma_{q,1}|^2  - |\Sigma_{q,1}|^2 \right]  \\
  && + \left(-\Lambda_{k;q} (\tau_0) \Lambda_{k;q}^{*} (\tau) + \Xi_{k;q} (\tau_0) \Xi_{k;q}^* (\tau) \right)|\Sigma_{k,2}|^2
 \left[ |\Gamma_{q,1}|^2  - |\Sigma_{q,1}|^2 \right] \ .
\end{eqnarray}
Since $|\Gamma_{k,i}|^2 -|\Sigma_{k,i}|^2 =1$ for any $k,i$, and considering the following realtions
\begin{eqnarray}\label{RelationbetweenBogoliubovCoeff}
 \Lambda_{k;q} (\tau_0) \Lambda_{k;q}^{*}(\tau)  =  \left(\Lambda_{k;q}^{*} (\tau_0) \Lambda_{k;q} (\tau) \right)^*\,, \qquad \Xi_{k;q} (\tau_0) \Xi_{k;q}^{*}(\tau) = \left(\Xi_{k;q}^{*} (\tau_0) \Xi_{k;q} (\tau) \right)^*\,,
\end{eqnarray}
we easily  verify Eq.(\ref{PartialInvariance})
which proves the invariance of the oscillation formulae.

\end{widetext}

\section{Flat spacetime limit}

The simplest application of the formalism, as well as a consistency check, is the flat spacetime limit. In flat space eq.\eqref{DefinitiveProbs}, reduces to the QFT formula derived in ref. \cite{BMix}.
Indeed, in this case, we can choose the Cauchy hypersurfaces to be the $t = constant$ surfaces in a given Minkowskian coordinate system, then the modes $\{\zeta_{\pmb{k},i} (x), \zeta^{*}_{\pmb{k},i}(x) \}$ coincide with the plane wave solutions of the flat Klein-Gordon equation, so that $\Lambda_{q;k} (t) \rightarrow U_{q;k}(t)$ and $\Xi_{q;k}(t) \rightarrow V_{q,r;k,s} (t)$. Here $U_{q;k}$ and $V_{q;k}$ are the Bogoliubov coefficients for boson mixing in flat space: $U_{q;k}=\delta^{3}(k-q)|U_{k}|\cdot e^{i(\omega_{k,2}-\omega_{k,1})t}$ and $V_{q;k}=\delta^{3}(k-q)|V_{k}|\cdot e^{i(\omega_{k,1}+\omega_{k,2})t}$. Here the label $k$ stands for the three-momentum $\pmb{k}$ and $\omega_{k,i} = \sqrt{k^2 + m_i^2}$.
The transition probability of eq.\eqref{DefinitiveProbs} is thus given by
\begin{widetext}
\begin{equation}\label{Probab}
\begin{split}
P_{A \rightarrow B}(t) &=2 \cos^{2}(\theta)\sin^{2}(\theta)(1-\Re[U^{*}_{k}(0) U_{k}(t)-V^{*}_{k}(0) V_{k}(t)]) \\
&=2 \cos^{2}(\theta) \sin^{2}(\theta)\bigg[1-\Re \left[|U_{\pmb{k}}(0)|^2 e^{i(\omega_{k,2} - \omega_{k,1})t} - |V_{\pmb{k}}(0)|^2e^{i(\omega_{k,2} + \omega_{k,1})t}\right]\bigg]\\
&=2 \cos^{2}(\theta) \sin^{2}(\theta)\bigg[1- |U_{\pmb{k}}(0)|^2 \cos [(\omega_{k,2} - \omega_{k,1})t] + |V_{\pmb{k}}(0)|^2 \cos [(\omega_{k,2} + \omega_{k,1})t]\bigg]\\
&=\sin^{2}(2\theta)\bigg[|U_{\pmb{k}}(0)|^2 \left(\frac{1}{2}-\cos[(\omega_{k,2} - \omega_{k,1})t] \right) +|V_{\pmb{k}}(0)|^2 \left(-\frac{1}{2}+\cos [(\omega_{k,2} + \omega_{k,1})t] \right)\bigg]\,,
\end{split}
\end{equation}
which reduces to the flat space oscillation formulas as expected
\begin{equation}\label{ProbRefound}
P_{A \rightarrow B}(t)=\sin^{2}(2\theta)\left[|U_{\pmb{k}}(0)|^2 \sin^{2}\left(\frac{\omega_{k,2} - \omega_{k,1}}{2}\right)t -|V_{\pmb{k}}(0)|^2\sin^{2}\left(\frac{\omega_{k,2} + \omega_{k,1}}{2}\right)t\right]\,.
\end{equation}
\end{widetext}

\section{Cosmological metrics}

As a non-trivial application let us consider the spatially flat Friedmann--Lemaitre--Robertson--Walker (FLRW) spacetimes, whose line element reads

\begin{equation}
ds^2 = dt^2 - a^2 (t) (dx^2 + dy^2 + dz^2) \ .
\end{equation}

For some choices of the scale factor $a(t)$ the Klein-Gordon equation \eqref{KGequation1} can be solved exactly. We will find it convenient to work with conformal time $d \eta = \frac{dt}{a(t)}$, in terms of which the line element becomes

\begin{equation}
 ds^2 = a^2 (\eta) \left( d \eta^2 - dx^2 - dy^2 - dz^2 \right) \ .
\end{equation}

\subsection{De Sitter expansion}

For a homogenous universe dominated by the cosmological constant the scale factor has an exponential evolution $a(t) = e^{H_0 t}$ with a constant Hubble expansion rate $H_0$. In conformal time $a (\eta) = \frac{-1}{H_0 \eta}$, with $\eta < 0$. We seek the basic plane wave solutions of the form $\zeta_{\pmb{k}} = (2\pi)^{-\frac{3}{2}}e^{i\pmb{k}\cdot \pmb{x}}a^{-1}(\eta)\chi_{k}(\eta)$, which inserted in the Klein-Gordon equation yields
\begin{equation}\label{BesselEquation}
\ddot{\chi}_{k}+\bigg(k^{2}+\frac{m^{2}}{H^{2}_{0}\eta^{2}}-\frac{2}{\eta^2}\bigg)\chi_{k}(\eta)=0 \  ,
\end{equation}
with $k = |\pmb{k}|$ and the dots denoting derivative with respect to $\eta$. This is a Bessel-like equation, and the general solution can be written as
\begin{equation}\label{BesselSolution}
\chi_{k}=(k\eta)^{\frac{1}{2}}(C_{1}H^{1}_{\nu}(k\eta)+C_{2}H^{2}_{\nu^{*}}(k\eta))
\end{equation}
with $\nu= \sqrt{\frac{9}{4}-\frac{m^{2}}{H^{2}_{0}}}$ and $H^{(m)}_{\nu}$ denoting Hankel functions of type $m$ and order $\nu$.
\begin{figure}
 \includegraphics[width = \linewidth]{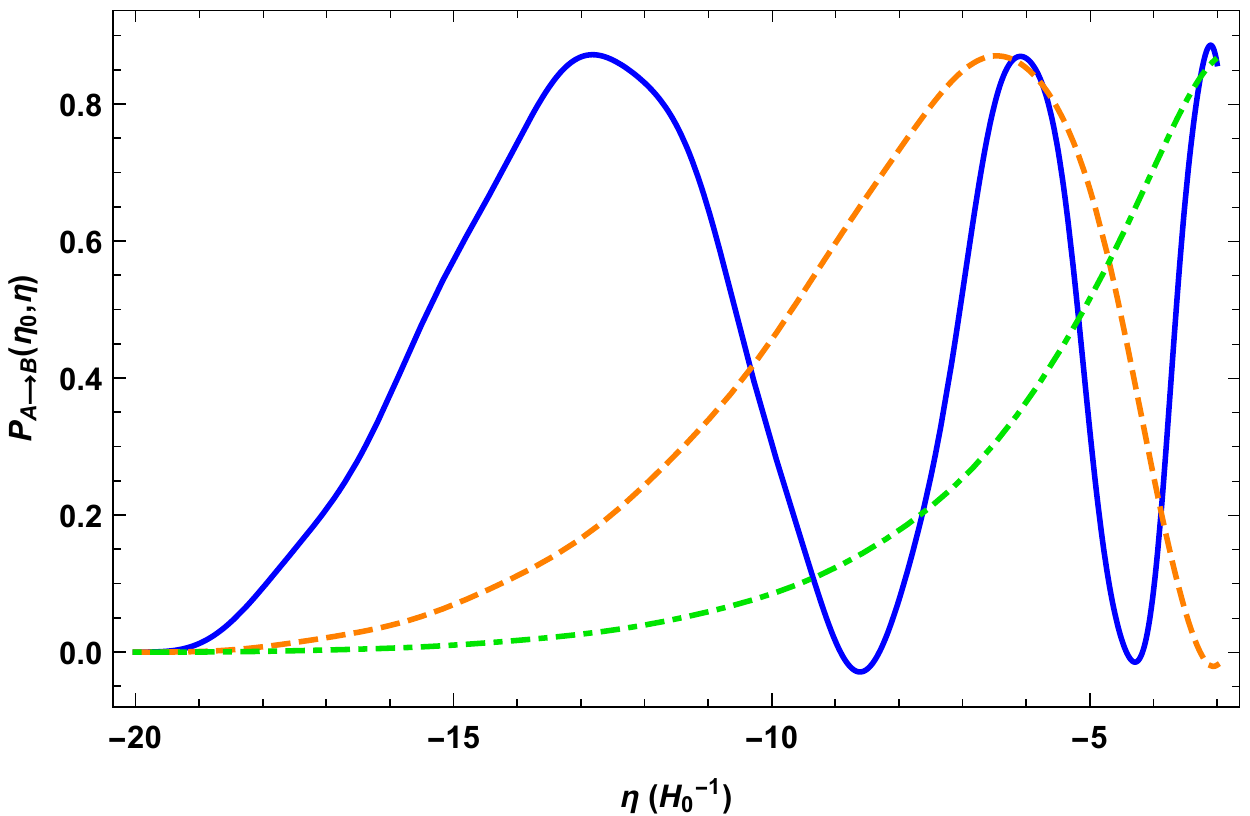}
 \caption{(color online): Plots of the oscillation formula for a universe with De Sitter expansion, as a function of conformal time $\eta$. Masses and momenta are expressed in units of $H_0$ and $\eta, \eta_0$ in units of $H_0^{-1}$. The blue solid line corresponds to $m_1 = 10$, $m_2 = 20$, $k=1$, the orange dashed line to $m_1=5$, $m_2 =10$, $k=1$ and the green dotdashed line to $m_1=3$, $m_2=6$ and $k=1$. The mixing angle is chosen as $\sin^2 2 \theta = 0.861$ and the reference time as $\eta_0 = - 20.1$. }
\end{figure}

To fix the boundary conditions we require that the modes $\zeta_{\pmb{k}}$ be positive with respect to $\partial_{\eta}$ at early times, i.e. $\zeta_{\pmb{k}} (\eta) \propto e^{-i k \eta}$ for $\eta \rightarrow - \infty$. This choice corresponds to the so-called adiabatic vacuum as the mass vacuum. In addition we impose the normalization condition $(\zeta_{\pmb{q}},\zeta_{\pmb{k}}) = \delta^{3} (\pmb{k}- \pmb{q})$.
The positive energy modes then read
\begin{equation}
\zeta_{\pmb{k}}=\bigg(\frac{-\pi H_0 k\eta^3}{4}\bigg)^{\frac{1}{2}}H^{1}_{\nu}(-k \eta) e^{i\frac{\pi}{4}(\nu-\nu^{*})} e^{i\pmb{k}\cdot \pmb{x}}
\end{equation}
where we recall that $\eta < 0$. Notice that for $m \leq \frac{3 H_0}{2}$, $\nu$ is real and the exponential factor $e^{i \frac{\pi}{4} \left( \nu - \nu^* \right)}$ equals $1$.

The Bogoliubov coefficients are now computed by taking the inner product of the modes corresponding to distinct masses. We have
\begin{widetext}
\begin{equation}\label{BCoefficient}
\begin{split}
\Lambda_{k,q}&=\delta^{3}(\pmb{k}-\pmb{q})i\frac{\pi k\eta}{4}(H^{1 *}_{\nu_{2}}\partial_{\eta}H^{1}_{\nu_{1}}- \partial_{\eta}H^{1*}_{\nu_{2}}H^{1}_{\nu_{1}})e^{i\frac{\pi}{4}(\nu_{1}-\nu_{1}^* + \nu_2 -\nu^{*}_{2})}\\
\Xi_{k,q}&=\delta^{3}(\pmb{k}-\pmb{q})i\frac{\pi k\eta}{4}(H^{2 *}_{\nu_{1}}\partial_{\eta}H^{2*}_{\nu_{2}}-\partial_{\eta}H^{2 *}_{\nu_{1}}H^{2*}_{\nu_{2}})e^{i\frac{\pi}{4}(\nu_{1}-\nu_{1}^* + \nu_2 -\nu^{*}_{2})}\, ,
\end{split}
\end{equation}
where we have omitted the argument $-k \eta$ of the Hankel functions. By introducing these expressions in \eqref{DefinitiveProbs}, we have the oscillation formula for the considered metric. The oscillation formulae are plotted for sample values of masses and momenta in figure (1).

\end{widetext}

\subsection{Universe dominated by radiation}

Let us now consider a scale factor for a universe dominated by radiation  $a(t) = a_{0}t^{\frac{1}{2}}$ and $a_{0}= constant$. In terms of the conformal time $\eta$, $a(\eta)$ is given by  $a(\eta)=\frac{a^{2}_{0}}{2}\eta$. Notice that the metric is defined only for positive $t$, which corresponds to positive conformal time $\eta = \frac{2 t^{\frac{1}{2}}}{a_0}$.
Employing the ansatz $\zeta_{\pmb{k}}=(2\pi)^{-\frac{3}{2}}e^{i\pmb{k} \cdot \pmb{x}}a^{-1}(\eta)\chi_{\pmb{k}}(\eta)$ in the Klein-Gordon equation, we obtain the equation
\begin{equation}\label{EquationforTemporalPart}
\ddot{\chi}_{k}+\bigg(k^{2}+\frac{m^{2}a^{4}_{0}}{4}\eta^{2}\bigg)\chi_{k}(\eta) \ .
\end{equation}

\begin{figure}
 \includegraphics[width = \linewidth]{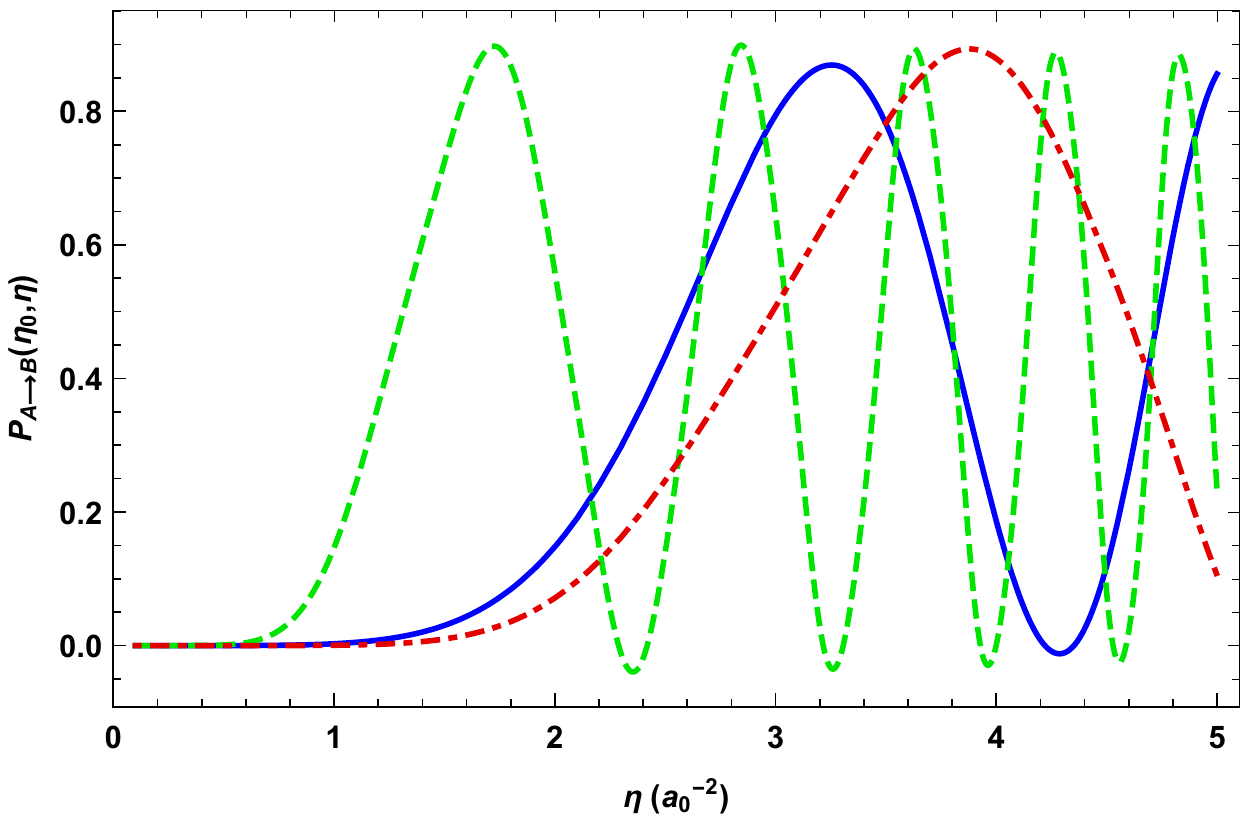}
 \caption{(color online): Plots of the oscillation formula for a universe dominated by radiation, as a function of conformal time $\eta$. Masses and momenta are expressed in units of $a_0^2$ and $\eta, \eta_0$ in units of $a_0^{-2}$. The blue solid line corresponds to $m_1 = 2$, $m_2 = 4$, $k=4$, the green dashed line to $m_1=5$, $m_2 =10$, $k=2$ and the red dotdashed line to $m_1=1$, $m_2=2$ and $k=1$. The mixing angle is chosen as $\sin^2 2 \theta = 0.861$ and the reference time as $\eta_0 = 0.1$. }
\end{figure}

It is convenient to introduce the dimensionless variable $\lambda = \sqrt{m} a_0 \eta$, in terms of which the equation reads

\begin{equation}
 \partial^2_{\lambda} \chi_{k} - \left(-\frac{k^2}{m a_0^2} - \frac{\lambda^2}{4} \right) \chi_{k} = 0
\end{equation}

This is the Weber equation: $\ddot{\chi}_{k}-(b+a\lambda^{2})\chi_{k}=0$,  with  $b=-\frac{k^{2}}{m a_0^2}$ and $a=-\frac{1}{4}$. It can be transformed in a degenerate hypergeometric equation by setting $z=\lambda^{2}\sqrt{a} = i \frac{\lambda^2}{2}$ and $\tilde{u}=\chi e^{\frac{\lambda}{2}}$:

\begin{equation}\label{HypergeometricEquation}
z\ddot{\tilde{u}}+\dot{\tilde{u}}\bigg(\frac{1}{2}-z\bigg)-\frac{1}{4}\bigg(-\frac{2k^{2}}{ima^{2}_{0}}\bigg)\tilde{u}=0\,.
\end{equation}

Therefore, the general solution can be written as:
\begin{widetext}
\begin{equation}\label{GeneralSolution}
\chi_{k}=e^{\frac{-ima^{2}_{0}\eta^{2}}{4}}\frac{1}{\sqrt{\eta}}\bigg(\frac{ima^{2}_{0}\eta^{2}}{2}\bigg)^{\frac{1}{4}}\left[C_{1}\phi\left(\frac{1}{4}\left(\frac{-2k^2}{ima^{2}_{0}}+1\right),\frac{1}{2},\frac{ima^{2}_{0}\eta^{2}}{2}\right)+C_{2}\psi \left(\frac{1}{4}\left(\frac{-2k^2}{ima^{2}_{0}}+1\right),\frac{1}{2},\frac{ima^{2}_{0}\eta^{2}}{2}\right)\right]
\end{equation}

Here we have explicitated $\lambda = \frac{i m a_0^2 \eta^2}{2}$. The functions $\phi (a, b, z)$ and $\psi (a,b,z)$ are the confluent hypergeometric functions (also denoted respectively as $M(a,b,z)$ and $U(a,b,z)$ \cite{Abramowitz}). At early times, we impose the normalization condition and we assume that the modes are positive with respect to $\partial_{\eta}$. We obtain
\begin{equation}\label{Solution}
\chi_{k}=\frac{(2\pi)^{\frac{3}{2}}}{\sqrt{4ma^{2}_{0}\eta}}e^{-\frac{\pi k^2}{4ma^{2}_{0}}}\bigg(\frac{ima^{2}_{0}\eta^{2}}{2}\bigg)^{\frac{1}{4}}e^{\frac{-ima^{2}_{0}\eta^{2}}{4}}
\psi\left(\frac{1}{4}\left(\frac{-2k^2}{ima^{2}_{0}}+1\right),\frac{1}{2},\frac{ima^{2}_{0}\eta^{2}}{2}\right)\,.
\end{equation}

Eq.(\ref{Solution}) can be  rewritten  in a more compact form by using  the relation between the hypergeometric functions and the Whittaker functions. We find
\begin{equation}\label{ModeZ}
\zeta_{k}=\sqrt{\frac{1}{m a_0^6 \eta^3}}
e^{-\frac{\pi k^2}{4 ma^{2}_{0}}}W_{-\frac{1}{4}\left(\frac{-2k^2}{ima^{2}_{0}}\right),\frac{1}{4}}\bigg(\frac{ima^{2}_{0}\eta^{2}}{2}\bigg) \ .
\end{equation}

where $W_{\kappa, \mu} (z)$ is the Whittaker $W$ function and we have omitted the spatial dependence.
The Bogoliubov coefficients are computed as usual and are:
\begin{equation}\label{Bcoeff}
\begin{split}
&\Lambda_{k,q}= \delta^{3}(k-q)\frac{i}{(a^{2}_{0}\eta)\sqrt{m_{1}m_{2}}}e^{-\frac{\pi k^2}{4a^{2}_{0}(m_{1}+m_{2})}} \times \\
&\bigg[W^{*}_{-\frac{1}{4}\left(\frac{-2k^2}{im_{2}a^{2}_{0}}\right),\frac{1}{4}}\left(\frac{im_{2}a^{2}_{0}\eta^{2}}{2}\right)\dot{W}_{-\frac{1}{4}\left(\frac{-2k^2}{im_{1}a^{2}_{0}}\right),\frac{1}{4}}\left(\frac{im_{1}a^{2}_{0}\eta^{2}}{2}\right)
-\dot{W}^{*}_{-\frac{1}{4}\left(\frac{-2k^2}{im_{2}a^{2}_{0}}\right),\frac{1}{4}}\left(\frac{im_{2}a^{2}_{0}\eta^{2}}{2}\right)W_{-\frac{1}{4}\left(\frac{-2k^2}{im_{1}a^{2}_{0}}\right),\frac{1}{4}}\left(\frac{im_{1}a^{2}_{0}\eta^{2}}{2}\right)\bigg]\\
&\Xi_{k,q}= \delta^{3}(k-q)\frac{i}{(a^{2}_{0}\eta)\sqrt{m_{1}m_{2}}}e^{-\frac{\pi k^2}{4a^{2}_{0}(m_{1}+m_{2})}} \times \\
&\bigg[W^{*}_{-\frac{1}{4}\left(\frac{-2k^2}{im_{1}a^{2}_{0}}\right),\frac{1}{4}}\left(\frac{im_{1}a^{2}_{0}\eta^{2}}{2}\right)\dot{W}^{*}_{-\frac{1}{4}\left(\frac{-2k^2}{im_{2}a^{2}_{0}}\right),\frac{1}{4}}\left(\frac{im_{2}a^{2}_{0}\eta^{2}}{2}\right) -\dot{W}^{*}_{-\frac{1}{4}\left(\frac{-2k^2}{im_{1}a^{2}_{0}}\right),\frac{1}{4}}\left(\frac{im_{1}a^{2}_{0}\eta^{2}}{2}\right)W^{*}_{-\frac{1}{4}\left(\frac{-2k^2}{im_{2}a^{2}_{0}}\right),\frac{1}{4}}\left(\frac{im_{2}a^{2}_{0}\eta^{2}}{2}\right)\bigg]
\end{split}
\end{equation}
Upon insertion in equation \eqref{DefinitiveProbs} one obtains the transition probabilities. The latter are plotted in figure (2) for sample values of masses and momenta.
\end{widetext}

\section{Conclusions}

We have constructed a curved space generalization of the QFT of boson mixing. We have derived new oscillation formulae and discussed the transformation properties of the theory under changes of mass representation. We have found that the interplay of curvature and mixing produces a rich structure which is considerably more complicated than that of the flat space theory. This is mirrored in a more involved form of the oscillation formulae, even for the comparatively simple spatially flat cosmological metrics considered in the paper.
The formalism has been developed under very general assumptions and may be applied, a priori, to any reasonable spacetime. Additional complications may be expected for all the known  concrete physical systems which exhibit boson mixing. Neutral meson mixing, for instance, concerns unstable particles, and the formalism here developed only applies as long as one can neglect the decay (that is, on sufficiently small timescales). Axion-photon mixing also presents specific difficulties due to the derivative form of the interaction Lagrangian responsible for mixing $L_{INT} \simeq g a F_{\mu \nu} \tilde{F}^{\mu \nu} $, in which the photon field clearly enters through derivative terms. A preliminary study on axion-photon mixing in QFT has been presented in Ref.\cite{Ax}.
The formalism may instead apply, without substantial modifications, to the mixing of hypothetical elementary bosons. That could be the case, for instance, of the supersymmetryc bosonic partners of neutrinos.
Yet the most interesting application is arguably the analysis of the corresponding bosonic flavor vacuum in curved space. From the preliminary results obtained in flat space \cite{FDM1,FDM2} one may expect the latter to yield a contribution to the dark energy of the universe. This aspect shall be carefully studied in a forthcoming paper.

\section*{Acknowledgements}
Partial financial support from MIUR and INFN is acknowledged.
A.C. also acknowledges the COST Action CA1511 Cosmology
and Astrophysics Network for Theoretical Advances and Training Actions (CANTATA).


\begin{thebibliography}{99}

\bibitem{Neutrino1}
S. M. Bilenky and B. Pontecorvo, Phys. Rep. \textbf{41.4}, pp. 225-261 (1978); S. M. Bilenky and S. T. Petcov, Rev. Mod. Phys. \textbf{59}, pp. 671-754 (1987); S. M. Bilenky, C. Giunti and W. Grimus, Progress in Particle and Nuclear Physics \textbf{43}, pp. 1-86 (1999); A. Capolupo, S. M. Giampaolo and A. Quaranta, Phys. Lett. B \textbf{820}, 136489 (2021); P. F. De Salas et al., JHEP \textbf{2021}, 71 (2021).

\bibitem{Neutrino2}
A. M. Gago, E. M. Santos, W. J. C. Teves and R. Zukanovich Funchal, Phys. Rev. D \textbf{63}, 073001 (2001); A. Capolupo, S. M. Giampaolo and G. Lambiase, Phys. Lett. B \textbf{792}, pp. 298-303 (2019); L. Buoninfante, A. Capolupo, S. M. Giampaolo
and G. Lambiase, Eur. Phys. J. C \textbf{80}, 1009 (2020); G. L. Fogli, E. Lisi, A. Marrone, D. Montanino and A. Palazzo, Phys. Rev. D \textbf{76}, 033006 (2007); A. Capolupo, S. M. Giampaolo, G. Lambiase and A. Quaranta, Universe \textbf{2020}, 6(11),
207 (2020).

\bibitem{FMix3}
E. Alfinito, M.Blasone, A.Iorio, G.Vitiello, Phys. Lett. B \textbf{362}, 91 (1995);

\bibitem{FMix1}
M. Blasone, A. Capolupo and G. Vitiello, Phys. Rev. D \textbf{66}, 025033 (2002) and references therein.

\bibitem{FMix2}
C.-R. Ji and Y. Mischchenko, Phys. Rev. D \textbf{64}, 076004 (2001).

\bibitem{Fmix4}
C.-R. Ji and Y. Mishchenko, Phys. Rev. D \textbf{65}, 096015 (2002).

\bibitem{Capolupo2020}
 A. Capolupo, G. Lambiase and A. Quaranta, Phys. Rev. D \textbf{101}, 095022 (2020).

\bibitem{FDM}
A. Capolupo, S. Carloni and A. Quaranta, Phys. Rev. D \textbf{105}, 105013 (2022).


\bibitem{FDM1}
A. Capolupo, Adv. High En. Phys. \textbf{2016}, 8089142 (2016).

\bibitem{FDM2}
A. Capolupo, Adv. High En. Phys. \textbf{2018}, 9840351 (2018).

\bibitem{FDM3}
A. Capolupo, S. Capozziello and G. Vitiello, Phys. Lett. A \textbf{373.6}, pp. 601-610 (2009).

\bibitem{FDM4}
A. Capolupo, S. Capozziello and G. Vitiello, Phys. Lett. A \textbf{363.1}, pp. 53-56 (2007).


\bibitem{Sim}
A. Capolupo, S. M. Giampaolo, G. Lambiase and A. Quaranta, Eur. Phys. J. C \textbf{80}, 423 (2020); A. Capolupo, S. M. Giampaolo and A. Quaranta, Eur. Phys. J. C \textbf{81}, 410 (2021).

\bibitem{Kabir}
P. K. Kabir, The CP puzzle, strange decays of the neutral kaon, Academic Press, (1968).

\bibitem{GellMann}
M. Gell-mann and A. Pais, Phys. Rev. \textbf{97}, 1387 (1955).

\bibitem{Jubb}
T. Jubb, M. Kirk, A. Lenz and G. Tetlalmatzi-Xolocotsi, Nucl. Phys. B \textbf{915}, pp. 431-453 (2017).

\bibitem{Bitenc}
U. Bitenc et al. (The Belle Collaboration), Phys. Rev. D \textbf{77}, 112003 (2008).

\bibitem{Meson1}
(LHCb collaboration), Nature Physics \textbf{18}, 1-5 (2022).

\bibitem{Meson2}
A. Lenz and U. Nierste, JHEP \textbf{06}, 072 (2007).

\bibitem{Meson3}
R. J. Dowdall et al., Phys. Rev. D \textbf{100}, 094508 (2019).

\bibitem{Meson4}
A. G. Grozin, T. Thomas and A. A. Pivovarov, Phys. Rev. D \textbf{98}, 054020 (2018).

\bibitem{Meson5}
A. Lenz et al., Phys. Rev. D \textbf{83}, 036004 (2011).

\bibitem{Meson6}
B. Aubert et al. (BABAR Collaboration), Phys. Rev. D \textbf{70}, 012007 (2004).

\bibitem{Meson7}
B. Aubert et al. (BABAR Collaboration), Phys. Rev. Lett. \textbf{100}, 131802 (2008).

\bibitem{Meson8}
A. Di Domenico (KLOE Collaboration), J. Phys. Conf. Ser. \textbf{171}, 012008 (2009).

\bibitem{Meson9}
B. Aubert et al. (BABAR Collaboration), Phys. Rev. Lett. \textbf{96}, 251802 (2006).

\bibitem{Meson10}
A. Di Domenico, Symmetry \textbf{2020}, 12 (12), 2063 (2020).

\bibitem{Meson11}
R. A. Bertlmann, W. Grimus and B. C. Hiesmayr, Phys. Rev. D \textbf{60}, 114032 (1999).



\bibitem{Axion1}
 G. Raffelt and L. Stodolsky, Phys. Rev. D \textbf{37}, pp. 1237-1249 (1988); P. W. Graham, I. G. Irastorza, S. K. Lamoreaux, A. Lindner and K. A. van Bibber, Annual Review of Nuclear and Particle Science \textbf{65}, pp. 485-514 (2015); A. Capolupo, G. Lambiase, A. Quaranta and S. M. Giampaolo \textbf{804}, 135407 (2020); A. K. Ganguly, P. Jain and S. Mandal, Phys. Rev. D \textbf{79}, 115014 (2009).

 \bibitem{Axion2}
 O. Mena, S. Razzaque and F. Villaescusa-Navarro, JCAP02 (2011) 030, (2011); A. Capolupo, S. M. Giampaolo and A. Quaranta, Eur. Phys. J. C \textbf{81}, 1116 (2021); Y. Grossman, S. Roy and J. Zupan, Phys. Lett. B \textbf{543}, Issues 1-2, pp. 23-28 (2002); D. J. E. Marsh, Phys. Rep. \textbf{643}, pp. 1-79 (2016).

\bibitem{sneutrino1}
Y. Grossman and Howard E. Haber, Phys. Rev. Lett. \textbf{78}, 3438 (1997).

\bibitem{sneutrino2}
S. Bar-Shalom, G. Eilam and A. Soni, Phys. Rev. Lett. \textbf{80}, 4629 (1998).

\bibitem{BMix}
M. Blasone, A. Capolupo, O. Romei and G. Vitiello, Phys. Rev. D \textbf{63}, 125015 (2001).

\bibitem{BMix2}
A. Capolupo, C.-R. Ji, Y. Mishchenko and G. Vitiello, Phys. Lett. B \textbf{594}, Issues 1-2, pp. 135-140 (2004);




\bibitem{Curv1}
 N. Birrell and P. Davies, Quantum Fields in Curved
Space (Cambridge Monographs on Mathematical Physics)
(Cambridge University Press, Cambridge, England, 1982).

\bibitem{Curv2}
R. Wald, Quantum Field Theory in Curved Spacetime
and Black Hole Thermodynamics (Chicago Lectures in
Physics) (The University of Chicago Press, Chicago, 1994),
ISBN: 9780226870274.

\bibitem{Curv3}
V. Mukhanov, S. Winitzki: Introduction to Quantum Effects in Gravity (Cambridge University Press, Cambridge, 2007).

\bibitem{Curv4}
L. Parker and D. Toms, Quantum Field Theory in Curved Spacetime (Cambridge University Press, Cambridge, 2009).

\bibitem{Abramowitz}
M. Abramowitz and I. A. Stegun, Handbook Of Mathematical Functions (Dover Publications Inc., New York, 1965),
ISBN-13: 978-0-486-61272-0.

\bibitem{Ax}
A. Capolupo, I. De Martino, G. Lambiase and An. Stabile,
%Axion-photon mixing in quantum field theory and vacuum energy, published on
Phys. Lett. B, \textbf{790}, 427  (2019).



% \bibitem{Garriga1991}
% J. Garriga and E. Verdaguer, Phys. Rev. D \textbf{43}, 2 (1991).

\end{thebibliography}
\end{document}